\documentclass[preprint,eqsecnum,aps,nofootinbib]{revtex4}
\usepackage[dvips]{graphicx}
\usepackage{amssymb}
\usepackage{amsmath}
\usepackage{amsmath, amsthm, amssymb, mathrsfs}

\begin{document}

\title{Three-Party Entanglement in Tripartite Teleportation 
                                                    Scheme through Noisy Channels}
\author{Eylee Jung, Mi-Ra Hwang,
DaeKil Park}

\affiliation{Department of Physics, Kyungnam University, Masan,
631-701, Korea}

\author{Sayatnova Tamaryan}

\affiliation{Theory Department, Yerevan Physics Institute,
Yerevan-36, 375036, Armenia}

\begin{abstract}
In this paper we have tried to interpret the physical role of the three-tangle and $\pi$-tangle
in the real physical information process. For the model calculation we adopt the 
three-party teleportation scheme through the various noisy channels. The three parties 
consist of sender, accomplice and receiver. It is shown that the $\pi$-tangles for the
X- and Z-noisy channels vanish at $\kappa t \rightarrow \infty$ limit, where $\kappa t$ is 
a parameter introduced in the master equation of Lindblad form. In this limit the receiver's 
maximum fidelity reduces to the classical limit $2/3$. However, this nice feature is not
maintained at the Y- and isotropy-noise channels. For Y-noise channel the $\pi$-tangle vanishes
at $0.61 \leq \kappa t$. 
At $\kappa t = 0.61$ the receiver's maximum fidelity becomes $0.57$, which
is much less than the classical limit. Similar phenomenon occurs at the isotropic noise channel.
We also computed the three-tangles analytically for the X- and Z-noise channels. The remarkable
fact is that the three-tangle for Z-noise channel is exactly same with the corresponding 
$\pi$-tangle. In the X-noise channel the three-tangle vanishes at $0.10 \leq \kappa t$. At 
$\kappa t = 0.10$ the receiver's fidelity can be reduced to the classical limit provided that 
the accomplice performs the measurement appropriately. However, the receiver's maximum
fidelity becomes $8/9$, which is much larger than the classical limit. Since the Y- and 
isotropy-noise channels are rank-$8$ mixed states, their three-tangles are not computed 
explicitly in this paper. Instead, we have derived their upper bounds with use of the 
analytical three-tangles for other noisy channels. Our analysis strongly suggests that 
we need different three-party entanglement measure whose value is between three-tangle 
and $\pi$-tangle.
\end{abstract}


\maketitle

\section{Introduction}

It is well-known that Entanglement of quantum state is a valuable physical 
resource in quantum information theories\cite{nielsen00}. It makes the quantum 
teleportation\cite{bennett93} and superdense coding\cite{bennett92} possible within the quantum 
mechanical law. Furthermore, it is the physical resource which is responsible for the 
speed-up of the quantum computer\cite{vidal03-1}. In this reason there has been a 
flurry of activity recently in the research of entanglement.

Many new properties for the entanglement of three- or higher-qubit pure states have been
reported in the recent papers\cite{pure}. However, it is in general much more difficult 
to understand the properties of entanglement for the mixed states except bipartite case.
These difficulties are mainly originated from the fact that the mixed state entanglement is 
defined by a convex-roof extension\cite{benn96,uhlmann99-1} of the pure state entanglement.
In order to compute the entanglement defined by convex-roof method one should derive the 
optimal decomposition for the given mixed state. Generally, however, it is highly non-trivial
task to derive the optimal decomposition for the arbitrary mixed states.  
This computational difficulty makes it difficult to characterize the multipartite entanglement
for the mixed states.

For the bipartite qubit states, fortunately, Wootters found how to derive the optimal 
decomposition for the concurrence, entanglement measure for the bipartite states, in
Ref.\cite{form2,form3}. Thus, one can compute the concurrence ${\cal C} (\rho)$ for the 
arbitrary mixed states $\rho$ by Wootters formula
\begin{equation}
\label{two4}
{\cal C} (\rho) = \mbox{max} (0, \lambda_1 - \lambda_2 - \lambda_3 - \lambda_4)
\end{equation}
where $\lambda_i$'s are the eigenvalues, in decreasing order, of the Hermitian matrix
$$
\sqrt{\sqrt{\rho} (\sigma_y \otimes \sigma_y) \rho^* (\sigma_y \otimes \sigma_y)
\sqrt{\rho}}. $$

Complete understanding on the bipartite concurrence makes it possible to derive the 
purely three-party entanglement, called three-tangle, for the three-qubit pure
states\cite{tangle1}. This arises from the observation that the three-qubit pure state
$|\psi_{ABC}\rangle$ satisfies the following inequality
\begin{equation}
\label{ckw1}
{\cal C}^2_{AB} + {\cal C}^2_{AC} \leq {\cal C}^2_{A (BC)}
\end{equation}
where
${\cal C}_{AB}$ and ${\cal C}_{AC}$ are concurrences for the reduced states 
$\rho^{AB} = \mbox{Tr}_C |\psi_{ABC}\rangle \langle \psi_{ABC}|$ and 
$\rho^{AC} = \mbox{Tr}_B |\psi_{ABC}\rangle \langle \psi_{ABC}|$, and ${\cal C}_{A (BC)}$ is 
a concurrence between a pair $BC$ and $A$. Therefore, ${\cal C}_{A (BC)}$ represents an 
total entanglement of the qubit $A$ arising due to the remaining qubits. For pure state
${\cal C}^2_{A (BC)}$ reduces to $4 \mbox{det} \rho^A$, where 
$\rho^A = \mbox{Tr}_{BC} |\psi_{ABC}\rangle \langle \psi_{ABC}|$ and is called one-tangle.
In this sense, the inequality (\ref{ckw1}) indicates that the total one-tangle is greater than 
sum of two-tangles. In addition, this observation naturally implies that 
$\tau_{ABC} \equiv {\cal C}^2_{A (BC)} - ({\cal C}^2_{AB} + {\cal C}^2_{AC})$, which is called
three-tangle, represents the purely three-way entanglement. For three-qubit pure state 
$|\psi\rangle = \sum_{i,j,k=0}^1 a_{ijk} |ijk\rangle$, the three-tangle $\tau_{ABC}$
becomes\cite{tangle1}
\begin{equation}
\label{3-tangle-1}
\tau_{ABC} = 4 |d_1 - 2 d_2 + 4 d_3|,
\end{equation}
where
\begin{eqnarray}
\label{3-tangle-2}
& &d_1 = a^2_{000} a^2_{111} + a^2_{001} a^2_{110} + a^2_{010} a^2_{101} + 
                                                              a^2_{100} a^2_{011
}
                                                              \\   \nonumber
& &d_2 = a_{000} a_{111} a_{011} a_{100} + a_{000} a_{111} a_{101} a_{010} + 
         a_{000} a_{111} a_{110} a_{001}
                                                              \\   \nonumber
& &\hspace{1.0cm} +
         a_{011} a_{100} a_{101} a_{010} + a_{011} a_{100} a_{110} a_{001} + 
         a_{101} a_{010} a_{110} a_{001}
                                                              \\   \nonumber
& &d_3 = a_{000} a_{110} a_{101} a_{011} + a_{111} a_{001} a_{010} a_{100}. 
\end{eqnarray}
The three-tangle defined by Eq.(\ref{3-tangle-1}) exactly coincides with the modulus of a 
Cayley's hyperdeterminant\cite{cay1845,miy03} and is an invariant quantity under the 
local $SL(2,\mathbb{C})$ transformation\cite{ver03,lei04}. 

The three-tangle (\ref{3-tangle-1}) has following two important properties. Firstly, for a 
completely separable ($A-B-C$) and biseparable ($A-BC$, $B-AC$, $AB-C$) states $\tau_{ABC}$
becomes zero. This means that the three-tangle is truly the pure three-party quantity related  
to the entanglement. Secondly, the three-tangles for the 
Greenberger-Horne-Zeilinger(GHZ)\cite{green89} and W\cite{dur00-1} states defined
\begin{equation}
\label{ghzandw}
|GHZ\rangle = \frac{1}{\sqrt{2}} \left( |000\rangle + |111\rangle \right) 
\hspace{1.0cm}
|W\rangle = \frac{1}{\sqrt{3}} \left( |001\rangle + |010\rangle + |100\rangle \right)
\end{equation}
become
\begin{equation}
\label{3-tangle-ghz-w}
\tau_{ABC} (|GHZ\rangle) = 1 \hspace{1.0cm} \tau_{ABC} (|W\rangle) = 0. 
\end{equation}
Since the whole three-qubit pure states can be classified by completely separable, 
biseparable, GHZ-type, and W-type states through stochastic local operation and classical 
communication(SLOCC)\cite{dur00-1}, Eq.(\ref{ghzandw}) indicates that the three-tangle does
not properly reflect the three-party entanglement of the W-type states.

For the mixed states the three-tangle is defined by a convex-roof 
method\cite{benn96,uhlmann99-1} as follows:
\begin{equation}
\label{mixed-3-tangle}
\tau_{ABC} (\rho) = \min \sum_i p_i \tau_{ABC} (\rho_i)
\end{equation}
where the minimum is taken over all possible ensembles of pure states. The pure state ensemble 
corresponding to the minimum $\tau_{ABC}$ is called optimal decomposition. It is in general 
highly difficult to derive the optimal decomposition for the arbitrary mixed states. 
Fortunately, Lohmayer et al\cite{tangle2} have derived recently the optimal decomposition
when the mixed state $\rho$ is 
\begin{equation}
\label{mixed-ghz-w}
\rho (p) = p |GHZ\rangle \langle GHZ| + (1-p) |W\rangle \langle W|
\end{equation}
and have computed the three-tangle explicitly. They also have found that the 
Coffman-Kundu-Wootters(CKW) inequality (\ref{ckw1}) holds for mixed states as well as 
pure states.
Subsequently, the three-tangle for the rank-$2$ mixed state composed of the generalized 
GHZ and generalized W states has been computed in Ref.\cite{tangle3}. In Ref.\cite{tangle4} 
furthermore, the optimal decompositions and the three-tangle for the rank-$3$ mixed state
composed of GHZ, W, and flipped W states are also explicitly derived. Most recently, the 
three-tangle for the rank-$4$ mixed states composed of $4$-different GHZ states are explicitly
computed in Ref.\cite{jung09-1}.

On the other hand, in order to reflect the three-party entanglement of the W-type states
properly we need to define new three-party entanglement measure different from the three-tangle.
One of the candidate is a $\pi$-tangle discussed in Ref.\cite{ou07-1}. The $\pi$-tangle is 
defined in terms of the global negativities\cite{vidal01-1} defined
\begin{equation}
\label{negativity-1}
{\cal N}^A = || \rho^{T_A} || - 1 \hspace{1.0cm}
{\cal N}^B = || \rho^{T_B} || - 1 \hspace{1.0cm}
{\cal N}^C = || \rho^{T_C} || - 1 
\end{equation}
where $||R|| = \mbox{Tr} \sqrt{R R^{\dagger}}$, and the superscripts $T_A$, $T_B$ and $T_C$
represent the partial transpositions for the $A$-qubit, $B$-qubit and $C$-qubit respectively.
Due to the separability criterion via partial transposition\cite{peres96,horod96,horod97} 
it is easy to show that the global negativities
vanish for the separable states. It is worthwhile noting that the computation of the global
negativities is relatively simple compared to concurrence or three-tangle for the mixed
states since it does not need the convex-roof extension. In addition, the negativities also
satisfy the monogamy inequality
\begin{equation}
\label{monogamy-2}
{\cal N}_{AB}^2 + {\cal N}_{AC}^2 \leq {\cal N}_{A(BC)}^2
\end{equation}
like concurrence. Then, the $\pi$-tangle is defined as 
\begin{equation}
\label{pi-1}
\pi_{ABC} = \frac{1}{3} (\pi_A + \pi_B + \pi_C )
\end{equation}
where
\begin{equation}
\label{pi-2}
\pi_A = {\cal N}_{A(BC)}^2 - ({\cal N}_{AB}^2 + {\cal N}_{AC}^2) \hspace{.5cm}
\pi_B = {\cal N}_{B(AC)}^2 - ({\cal N}_{AB}^2 + {\cal N}_{BC}^2) \hspace{.5cm}
\pi_C = {\cal N}_{(AB)C}^2 - ({\cal N}_{AC}^2 + {\cal N}_{BC}^2).
\end{equation} 
It is easy to show that the $\pi$-tangles for $|GHZ\rangle$ and $|W\rangle$ become
\begin{equation}
\label{pi-ghz-w}
\pi_{ABC} (|GHZ\rangle) = 1 \hspace{1.0cm}
\pi_{ABC} (|W\rangle) = \frac{4}{9} (\sqrt{5} - 1) \sim 0.55.
\end{equation}
Thus the $\pi$-tangle reflects the three-party entanglement of the W-type states unlike the 
three-tangle.

In this paper we would like to explore the physical role of the three-party entanglement in the 
real quantum information process. In order to discuss this issue we adopt the 
tripartite teleportation scheme discussed in Ref.\cite{karl98}. Similar issue was discussed 
in Ref.\cite{08-mixed}, where the physical role of the concurrence is discussed in the 
bipartite teleportation through noisy channels. Ref.\cite{08-mixed} has shown that the
concurrences of the mixed state quantum channels arising due to some noises vanish in the 
region of $\bar{F} \leq 2 / 3$, where $\bar{F}$ is an average fidelity between initial 
Alice's unknown state and final Bob's state. Since $\bar{F} = 2/3$ corresponds to the best
possible score when Alice and Bob communicate with each other through the classical 
channel\cite{popescu94-1}, this result indicates that the entanglement of the quantum 
channel is a genuine physical resource for the teleportation process.

This paper is organized as follows. In section II we re-formulate the tripartite teleportation 
process\cite{karl98} in terms of the density matrix. This re-description allows us to 
formulate the tripartite teleportation process when quantum channel is mixed state. The
several basic quantities are calculated in this section, which are essential for the 
calculation of various fidelities in next sections. In section III we compute the accomplice's 
fidelities and receiver's fidelities when the tripartite teleportation process is performed
through noisy channels. 
In section IV we compute the $\pi$-tangles for the various noisy channels. 
It is shown that the $\pi$-tangles for all noise channels decrease with increasing 
the decoherence parameter $\kappa t$. This is in fact expected due to the fact that 
the decoherence in general disentangles the entanglement of quantum states like
``sudden death''. The $\pi$-tangle for X- and Z-noise channels vanish at the 
$\kappa t \rightarrow \infty$ limit. However, the $\pi$-tangles for Y- and isotropy-noise
channels are found to be non-zero at the finite range of $\kappa t$.
In section V we compute the three-tangles for the X- and Z-noise channels. It is shown that 
the three-tangle for the Z-noise channel is exactly same with the corresponding $\pi$-tangle.
The three-tangle for the X-noise channel is shown to have three different expressions depending
on the range of $\kappa t$. Since the channels for the Y- and isotropy-noises are 
rank-$8$ mixed states, there is no general method to compute the three-tangles. However, we 
derived the upper bound of these three-tangles.
In section VI we analyze the $\pi$-tangle and three-tangle by making use of the receiver's
fidelities. The $\pi$-tangle seems to be too large to have a nice physical interpretation. 
The three-tangle also seems to be too small by similar manner. This analysis strongly
suggests that we may need different three-party entanglement measure whose value is between 
three-tangle and $\pi$-tangle.

\section{Basic Quantities}

In this section we want to re-formulate the tripartite teleportation scheme in terms of the 
density matrices\cite{yeo03-1}. It involves sender (Alice), accomplice (Bob) and  
receiver (Charlie). Initially they share each single qubit of the GHZ state, {\it i.e.}
$\rho_{GHZ} = |GHZ\rangle_{234} \langle GHZ|$. The purpose of the tripartite teleportation is 
as follows. Firstly, Alice at location $2$ should transport a single qubit state
\begin{equation}
\label{unknown-1}
\rho_{in} = |\psi_{in}\rangle \langle \psi_{in}| \hspace{1.0cm}
|\psi_{in}\rangle = \cos \left(\frac{\theta}{2}\right) e^{i \phi/2} |0\rangle + 
                    \sin \left(\frac{\theta}{2}\right) e^{-i \phi/2} |1\rangle 
\end{equation}
to the receiver, Charlie, at location $4$ with fidelity $\bar{F}_C$ as high as possible  
with the help of the accomplice, Bob, at location $3$. At the second time Alice should
transport $\rho_{in}$ to the accomplice, Bob, with fidelity $\bar{F}_B$ as high as 
possible. Of course, we cannot make $\bar{F}_B = \bar{F}_C = 1$ due to 
no-cloning/broadcast theorems\cite{wootters82,barnum96}. The task is accomplished if one 
can make $\bar{F}_B$ and $\bar{F}_C$ as high as possible. In this sense the tripartite 
teleportation scheme is similar to a quantum copier 
(cloning device)\cite{buzek96-1,buzek97-1,buzek97-2,gisin97-1}.

From the postulate of quantum mechanics on composite systems the initial state of the 
tripartite teleportation process should be
\begin{equation}
\label{initial-1}
\rho_{in} \otimes \rho_{GHZ}.
\end{equation}
As will be discussed below $\rho_{GHZ}$ will be changed into $\varepsilon (\rho_{GHZ})$
when noise is introduced when Alice, Bob and Charlie prepare the GHZ state initially, 
where $\varepsilon$ is a quantum operation\cite{nielsen00}.

At the next stage Alice performs a projective measurement by preparing a set of the 
measurement operators $\{M_1, M_2, M_3, M_4 \}$ with
\begin{equation}
\label{projective-1}
M_1 = |\Phi^+\rangle \langle \Phi^+|    \hspace{.5cm}
M_2 = |\Phi^-\rangle \langle \Phi^-|    \hspace{.5cm}
M_3 = |\Psi^+\rangle \langle \Psi^+|    \hspace{.5cm}
M_4 = |\Psi^-\rangle \langle \Psi^-|,
\end{equation}
where
\begin{equation}
\label{bell-basis}
|\Phi^{\pm}\rangle = \frac{1}{\sqrt{2}} \left( |00\rangle \pm |11\rangle \right)_{12}
\hspace{1.0cm}
|\Psi^{\pm}\rangle = \frac{1}{\sqrt{2}} \left( |01\rangle \pm |10\rangle \right)_{12}.   
\end{equation}
Since $|\Phi^{\pm}\rangle$ and $|\Psi^{\pm}\rangle$ form a Bell basis, the operators satisfy 
the completeness constraint
\begin{equation}
\label{complete-1}
\sum_m M_m^{\dagger} M_m = I.
\end{equation}
From the quantum mechanical postulates the probability $P_{m}$, probability that the result
of the Alice's measurement is $m$, is given by
\begin{equation}
\label{probability-1}
P_m = \mbox{Tr} \left[ \left(M_m^{\dagger} M_m \otimes I_{34} \right) 
                       \left( \rho_{in} \otimes \rho_{GHZ} \right) \right]
\end{equation}
and the state of the system after the Alice's measurement reduces to
\begin{equation}
\label{post-1}
\tilde{\rho}_m = \frac{1}{P_m} (M_m \otimes I_{34}) \left( \rho_{in} \otimes \rho_{GHZ} \right)
                    (M_m \otimes I_{34})^{\dagger}.
\end{equation}
For our case we have $P_1 = P_2 = P_3 = P_4 = 1/4$. After measurement, Alice broadcasts her
measurement outcome to Bob and Charlie via a classical channel. 

Next, our concern is moved to the subsystem of Bob and Charlie. This process can be performed 
by tracing out the Alice's subsystem, {\it i.e}
\begin{equation}
\label{partial-1}
\pi_{3,4}^m = \mbox{Tr}_{1,2} \left(\tilde{\rho}_m \right).
\end{equation}
After then, the accomplice, Bob, performs a projective measurement again by preparing a set of 
measurement operators $\{N_1, N_2 \}$ with
\begin{equation}
\label{projective-2}
N_1 = |\mu^+\rangle \langle \mu^+| \hspace{1.0cm}
N_2 = |\mu^-\rangle \langle \mu^-|,
\end{equation}
where
\begin{equation}
\label{basis-2}
|\mu^+\rangle = \sin \nu |0\rangle + \cos \nu |1\rangle     \hspace{1.0cm}
|\mu^-\rangle = \cos \nu |0\rangle - \sin \nu |1\rangle.
\end{equation}
Since $|\mu^+\rangle$ and $|\mu^-\rangle$ form a basis for the Bob's qubit, the completeness
condition
\begin{equation}
\label{complete-2}
N_1^{\dagger} N_1 + N_2^{\dagger} N_2 = I
\end{equation}
is naturally satisfied. From the quantum mechanical postulates again the probability $q_{mn}$,
probability that the result of the Bob's measurement is $n$ on condition that the outcome of
Alice's measurement is $m$, reduces to
\begin{equation}
\label{probability-2}
q_{mn} = \mbox{Tr} \left[ (N_n \otimes I_4) \pi_{3,4}^m \right]
\end{equation}
and the state of the system after the Bob's measurement becomes
\begin{equation}
\label{post-2}
\tilde{\pi}_{mn} = \frac{1}{q_{mn}} (N_n \otimes I_4) \pi_{3,4}^m (N_n \otimes I_4)^{\dagger}.
\end{equation}
For our case $q_{mn}$ becomes
\begin{equation}
\label{probability-3}
q_{11} = q_{21} = q_{32} = q_{42} = \frac{1}{2} \left(1 - \cos 2 \nu \cos \theta \right)
\hspace{.5cm}
q_{12} = q_{22} = q_{31} = q_{41} = \frac{1}{2} \left(1 + \cos 2 \nu \cos \theta \right).
\end{equation}

After then, our concern is moved to the subsystem of Charlie by tracing out the Bob's 
subsystem, {\it i.e.}
\begin{equation}
\label{partial-2}
\chi_4^{mn} = \mbox{Tr}_3 \left( \tilde{\pi}_{mn} \right).
\end{equation}
Finally, Charlie takes an appropriate unitary transformation to his own qubit
\begin{equation}
\label{unit-1}
\tau_{mn} = u_4^{mn} \chi_4^{mn} \left(u_4^{mn} \right)^{\dagger}.
\end{equation}
The unitary operator $u_4^{mn}$ becomes 
\begin{equation}
\label{unit-2}
u_4^{11} = u_4^{22} = I  \hspace{.5cm}
u_4^{12} = u_4^{21} = \sigma_z  \hspace{.5cm}
u_4^{31} = u_4^{42} = \sigma_x  \hspace{.5cm}
u_4^{32} = u_4^{41} = \sigma_y  
\end{equation}
where $\sigma_i$ is usual Pauli matrices. At this stage the tripartite teleportation process
is terminated.

Now, we want to discuss the tripartite teleportation process through a noisy channel.
If noise is introduced at the initial stage when Alice, Bob, and Charlie share their each
single qubit of $|GHZ\rangle$, $\rho_{GHZ}$ is, in general, changed into the mixed state.
The mixed state can be derived by solving a master equation in the Lindblad 
form\cite{lindblad76}
\begin{equation}
\label{lindblad}
\frac{\partial \rho}{\partial t} = -i [H_S, \rho] + \sum_{i, \alpha}
\left(L_{i,\alpha} \rho L_{i,\alpha}^{\dagger} - \frac{1}{2}
\left\{ L_{i,\alpha}^{\dagger} L_{i,\alpha}, \rho \right\} \right)
\end{equation}
where the Lindblad operator 
$L_{i,\alpha} \equiv \sqrt{\kappa_{i,\alpha}} \sigma^{(i)}_{\alpha}$ acts on the 
$i$th qubit and describes decoherence. Of course, the operator
$\sigma^{(i)}_{\alpha}$ denotes the 
Pauli matrix of the $i$th qubit with $\alpha = x,y,z$. The constant $\kappa_{i,\alpha}$
is approximately equal to the inverse of decoherence time. In this paper we will assume for 
simplicity that the constant $\kappa_{i,\alpha}$ is independent of $i$ and $\alpha$, {\it i.e.}
$\kappa_{i,\alpha} = \kappa$.

Solutions of Eq.(\ref{lindblad}) for the 
$(L_{2,x}, L_{3,x}, L_{4,x})$, $(L_{2,y}, L_{3,y}, L_{4,y})$,
$(L_{2,z}, L_{3,z}, L_{4,z})$, and isotropy noises were solved explicitly in Ref.\cite{all1}.
The spectral decompositions of the results are as follows:
\begin{eqnarray}
\label{summary1}
& &\varepsilon_X (\rho_{GHZ}) = x |GHZ, 1\rangle \langle GHZ, 1| + 
\frac{1-x}{3} \bigg[|GHZ, 3\rangle \langle GHZ, 3| 
                                                           \\   \nonumber
& & \hspace{2.0cm} + |GHZ, 5\rangle \langle GHZ, 5| + 
|GHZ, 7\rangle \langle GHZ, 7| \bigg]  \hspace{1.0cm}  
\left( x = \frac{1}{4} (1 + 3 e^{-4 \kappa t}) \right)   \\  \nonumber
& &\varepsilon_Y (\rho_{GHZ}) = \frac{y_+^3}{8} |GHZ, 1\rangle \langle GHZ, 1| + 
\frac{y_-^3}{8} |GHZ, 2\rangle \langle GHZ, 2|           \\  \nonumber
& & \hspace{2.0cm} + \frac{y_+ y_-^2}{8} \bigg[ |GHZ, 3\rangle \langle GHZ, 3| 
+ |GHZ, 5\rangle \langle GHZ, 5| + 
|GHZ, 7\rangle \langle GHZ, 7| \bigg]                  \\   \nonumber 
& & \hspace{2.0cm} + \frac{y_+^2 y_-}{8} \bigg[ |GHZ, 4\rangle \langle GHZ, 4|
+ |GHZ, 6\rangle \langle GHZ, 6| +
|GHZ, 8\rangle \langle GHZ, 8| \bigg]                  \\   \nonumber
& & \hspace{11.0cm} (y_{\pm} = 1 \pm e^{-2 \kappa t})     \\   \nonumber
& &\varepsilon_Z (\rho_{GHZ}) = z |GHZ, 1\rangle \langle GHZ, 1| +  
(1-z) |GHZ, 2\rangle \langle GHZ, 2|  \hspace{1.0cm} \left(z = \frac{1}{2} (1 + e^{-6 \kappa t})
                                                                      \right)  
                                                                             \\   \nonumber
& &\varepsilon_I (\rho_{GHZ}) = \frac{1 + 3 p^2 + 4 p^3}{8} |GHZ, 1\rangle \langle GHZ, 1|
+  \frac{1 + 3 p^2 - 4 p^3}{8} |GHZ, 2\rangle \langle GHZ, 2|          
                                                                            \\  \nonumber
& & \hspace{3.0cm} + \frac{1 - p^2}{8} \bigg[ I - (|000\rangle \langle000| 
                                            + |111\rangle \langle111| ) \bigg]
\hspace{1.0cm} (p = e^{-4 \kappa t})
\end{eqnarray}
where the subscripts $X$, $Y$, $Z$, and $I$ represent the type of noise channels, and 
\begin{eqnarray}
\label{summary2}
& &|GHZ, 1\rangle = \frac{1}{\sqrt{2}} \left( |000\rangle + |111\rangle \right) \hspace{1.0cm}
|GHZ, 2\rangle = \frac{1}{\sqrt{2}} \left( |000\rangle - |111\rangle \right)
                                                                           \\   \nonumber
& &|GHZ, 3\rangle = \frac{1}{\sqrt{2}} \left( |001\rangle + |110\rangle \right) \hspace{1.0cm}
|GHZ, 4\rangle = \frac{1}{\sqrt{2}} \left( |001\rangle - |110\rangle \right)
                                                                           \\   \nonumber
& &|GHZ, 5\rangle = \frac{1}{\sqrt{2}} \left( |010\rangle + |101\rangle \right) \hspace{1.0cm}
|GHZ, 6\rangle = \frac{1}{\sqrt{2}} \left( |010\rangle - |101\rangle \right)
                                                                           \\   \nonumber
& &|GHZ, 7\rangle = \frac{1}{\sqrt{2}} \left( |011\rangle + |100\rangle \right) \hspace{1.0cm}
|GHZ, 8\rangle = \frac{1}{\sqrt{2}} \left( |011\rangle - |100\rangle \right).
\end{eqnarray}

\begin{center}
\begin{tabular}{c||c|c|c} \hline
quantities & no noise and $Z$ noise & $X$ and $Y$ noises  & Isotropy noise  \\  \hline \hline
$P_1, P_2, P_3, P_4$ & $\frac{1}{4}$ & $\frac{1}{4}$ & $\frac{1}{4}$     \\    \hline  
$q_{11}, q_{21}, q_{32}, q_{42}$ & $\frac{1}{2} (1 - \cos 2 \nu \cos \theta)$ & 
$\frac{1}{2} (1 - \cos 2 \nu \cos \theta e^{-4 \kappa t})$ & 
$\frac{1}{2} (1 - \cos 2 \nu \cos \theta e^{-8 \kappa t})$   \\   \hline
$q_{31}, q_{41}, q_{12}, q_{22}$ & $\frac{1}{2} (1 + \cos 2 \nu \cos \theta)$ & 
$\frac{1}{2} (1 + \cos 2 \nu \cos \theta e^{-4 \kappa t})$ &   
$\frac{1}{2} (1 + \cos 2 \nu \cos \theta e^{-8 \kappa t})$   \\   \hline
\end{tabular}

\vspace{0.1cm}
Table I: Basic Quantities in Tripartite Teleportation
\end{center}
\vspace{0.5cm}
The probabilities $P_m$'s and $q_{mn}$'s in the noisy channels can be directly computed 
by changing $\rho_{GHZ}$ into the mixed states (\ref{summary1}) in Eq.(\ref{probability-1})
and Eq.(\ref{probability-2}). The results for the $(L_{2,x}, L_{3,x}, L_{4,x})$, 
$(L_{2,y}, L_{3,y}, L_{4,y})$, $(L_{2,z}, L_{3,z}, L_{4,z})$\footnote{For simplicity, we will
use the terminology X-, Y-, and Z-noises together for $(L_{2,x}, L_{3,x}, L_{4,x})$,
$(L_{2,y}, L_{3,y}, L_{4,y})$, $(L_{2,z}, L_{3,z}, L_{4,z})$ noises},
and isotropy noise channels
are summarized in Table I. As Table I indicated, $P_m$'s and $q_{mn}$'s in the various 
noisy channels reduce to $P_i = 1/4 (i=1, \cdots, 4)$ and Eq.(\ref{probability-3}) when 
$\kappa = 0$ limit.

\section{Fidelities}

\begin{center}
\begin{tabular}{c||c|c} \hline
Type of noise &  $F_C (\theta, \phi)$ & $\bar{F}_C$ 
                                                                      \\   \hline \hline
no noise & $1 - \frac{1}{2} (1 - \sin 2 \nu ) \sin^2 \theta$ & $\frac{1}{3} (2 + \sin 2 \nu)$ 
                                                            \\ \hline
X noise & $\frac{1}{2} [(1 + \sin^2 \theta \cos^2 \phi \sin 2 \nu)$
& $\frac{1}{6} [ (3 + \sin 2 \nu)$   \\
{}  & $ + e^{-4 \kappa t} (\cos^2 \theta + \sin^2 \theta \sin^2 \phi \sin 2 \nu) ]$
& $+ e^{-4 \kappa t} (1 + \sin 2 \nu) ]$ \\  \hline
Y noise & $\frac{1}{2} [1 + e^{-2 \kappa t} \sin^2 \theta \sin^2 \phi \sin 2 \nu + 
e^{-4 \kappa t} \cos^2 \theta$
& $\frac{1}{6} [3 + e^{-2 \kappa t} \sin 2 \nu$   \\
{} & $+ e^{-6 \kappa t} \sin^2 \theta \cos^2 \phi \sin 2 \nu]$
& $ + e^{-4 \kappa t} + e^{-6 \kappa t} \sin 2 \nu]$      \\   \hline
Z noise & $1 - \frac{1}{2} (1 - \sin 2 \nu e^{-6 \kappa t}) \sin^2 \theta$
& $\frac{1}{3} [2 + e^{-6 \kappa t} \sin 2 \nu]$           \\   \hline
Isotropy noise & $\frac{1}{2} [1 + e^{-8 \kappa t} \cos^2 \theta + e^{-12 \kappa t} 
\sin^2 \theta \sin 2 \nu]$  &
$\frac{1}{6} [3 + e^{-8 \kappa t} + 2 \sin 2 \nu e^{-12 \kappa t}]$   \\   \hline 

\end{tabular}

\vspace{0.1cm}
Table II: Charlie's fidelity 
\end{center}
\vspace{0.5cm}

The Charlie's fidelity, which measures how well the initial state $\rho_{in}$ is transported
to the Charlie's final state, can be computed as follows. Since Charlie's final state is 
$\tau_{mn}$ provided that Alice and Bob measure $m$ and $n$ respectively, one can define
the fidelity $F_{mn}^C$ in this case as a form
\begin{equation}
\label{fidel-C-1}
F_{mn}^C = \mbox{Tr} \left[ \tau_{mn} \rho_{in} \right].
\end{equation}
Averaging over all possible measurement outcomes, we can define the Charlie's fidelity in a 
form
\begin{equation}
\label{fidel-C-2}
F_C (\theta, \phi) = \sum_{m=1}^{4} \sum_{n=1}^2 P_{m} q_{mn} F_{mn}^C.
\end{equation}
Finally, averaging $F_C (\theta, \phi)$ over all possible input states, we can define the 
Charlie's average fidelity $\bar{F}_C$ as follows:
\begin{equation}
\label{fidel-C-3}
\bar{F}_C = \frac{1}{4 \pi} \int_0^{\pi} d\theta \int_0^{2 \pi} d\phi \sin \theta 
F_C (\theta, \phi).
\end{equation}
When there is no noise, $F_C (\theta, \phi)$ and $\bar{F}_C$ becomes
\begin{equation}
\label{fidel-C-4}
F_C (\theta, \phi) = 1 - \frac{1}{2} (1 - \sin 2 \nu) \sin^2 \theta   \hspace{1.0cm}
\bar{F}_C = \frac{1}{3} (2 + \sin 2 \nu).
\end{equation}
Thus, the Charlie's fidelities depend on the set of Bob's measurement operators. If Bob 
chooses $\nu = \pi / 4$, $\bar{F}_C$ reaches to its maximum $\bar{F}_C = 1$, which means the
perfect teleportation from Alice to Charlie. 

The Charlie's fidelities $F_C (\theta, \phi)$ and $\bar{F}_C$ are summarized in Table II when 
the mixed states changed from $|GHZ\rangle$ by various noises are introduced 
as a quantum channel. Comparing Table II with Table I of 
Ref.\cite{all1}, one can realize that the Charlie's fidelities with $\nu = \pi/4$ exactly 
coincide with fidelities of the bipartite teleportation when same noises are introduced 
initially in the quantum channel.

In the tripartite teleportation scheme, however, there are additional fidelities 
between Alice's state $\rho_{in}$ and Bob's final state. 
Since Bob's final state after his measurement is 
$N_1$ or $N_2$ defined in Eq.(\ref{projective-2}) with respective probability
$\sum_{i=1}^4 P_i q_{i1}$ or $\sum_{i=1}^4 P_i q_{i2}$, the Bob's final fidelities
can be defined as 
\begin{eqnarray}
\label{fidel-B-1}
& &F_B^T (\theta, \phi) = \mbox{Tr} [N_1 \rho_{in}] \sum_{i=1}^4 P_i q_{i1} + 
                        \mbox{Tr} [N_2 \rho_{in}] \sum_{i=1}^4 P_i q_{i2}  \\  \nonumber
& &\bar{F}_B^T = \frac{1}{4 \pi} \int_0^{\pi} d\theta \int_0^{2\pi} d\phi \sin \theta
                                     F_B^T (\theta, \phi).
\end{eqnarray}
If one computes $F_B^T (\theta, \phi)$ and $\bar{F}_B^T$ for X-, Y-, Z-, and 
isotropy-noise channels, one can show that they are all same as
\begin{equation}
\label{fidel-B-2}
F_B^T (\theta, \phi) = \bar{F}_B^T = \frac{1}{2}.
\end{equation}
This is too small because the optimal value for a classical teleportation scheme is 
$2/3$.

\begin{center}
\begin{tabular}{c||c|c|c} \hline
Type of noise &  Alice's outcome & $F_B^m (\theta, \phi)$ & $\bar{F}_B^m$
                                                                      \\   \hline \hline
X-noise & $m=1, 2$ & $\frac{1}{2} + \frac{1}{2} \cos 2 \nu \cos 2 \theta$ &
$\frac{1}{6} \left(3 + e^{-4 \kappa t} \cos^2 2 \nu \right)$             \\ 
and & {} & $\times \left(\cos 2 \nu \cos \theta - \sin 2 \nu \sin \theta \right)
                         e^{-4 \kappa t}$ & {}                      \\   \cline{2-4}
Y-noise & $m=3, 4$ & $\frac{1}{2} - \frac{1}{2} \cos 2 \nu \cos 2 \theta$ & 
$\frac{1}{6} \left(3 - e^{-4 \kappa t} \cos^2 2 \nu \right)$     \\
{} & {} & $\times \left(\cos 2 \nu \cos \theta - \sin 2 \nu \sin \theta \right)
                         e^{-4 \kappa t}$ & {}                   \\   \hline
no-noise & $m=1, 2$ & $\frac{1}{2} \bigg( 1 + \cos^2 2 \nu \cos^2 \theta $ &
$\frac{2}{3} - \frac{1}{6} \sin^2 2 \nu$                          \\ 
and & {} & $ - \sin 2 \nu \cos 2 \nu \sin \theta \cos \theta \cos \phi \bigg)$ & {}
                                                                         \\   \cline{2-4}
Z-noise & $m=3, 4$ & $\frac{1}{2} \bigg( 1 - \cos^2 2 \nu \cos^2 \theta$ & 
$\frac{1}{3} + \frac{1}{6} \sin^2 2 \nu$     \\
{} & {} & $+ \sin 2 \nu \cos 2 \nu \sin \theta \cos \theta \cos \phi \bigg)$ & {}
                                                                            \\   \hline
{} & $m=1, 2$ & $\frac{1}{2} + \frac{1}{2} \cos 2 \nu \cos 2 \theta$ &
$\frac{1}{6} \left(3 + e^{-8 \kappa t} \cos^2 2 \nu \right)$             \\ 
Isotropy & {} & $\times \left(\cos 2 \nu \cos \theta - \sin 2 \nu \sin \theta \right)
                         e^{-8 \kappa t}$ & {}                      \\   \cline{2-4}
noise & $m=3, 4$ & $\frac{1}{2} - \frac{1}{2} \cos 2 \nu \cos 2 \theta$ & 
$\frac{1}{6} \left(3 - e^{-8 \kappa t} \cos^2 2 \nu \right)$     \\
{} & {} & $\times \left(\cos 2 \nu \cos \theta - \sin 2 \nu \sin \theta \right)
                         e^{-8 \kappa t}$ & {}                   \\   \hline
\end{tabular}

\vspace{0.1cm}
Table III: Bob's fidelities just after Alice broadcasts her outcome.
\end{center}
\vspace{0.5cm}

However, one can define the Bob's fidelities at the stage just after Alice broadcasts her 
measurement outcome to Bob and Charlie via classical channel. If Alice's outcome is $m$, 
then the 
Bob's fidelities can be defined as 
\begin{eqnarray}
\label{fidel-B-3}
& &F_{B}^m (\theta, \phi) = q_{m1} \mbox{Tr} [N_1 \rho_{in}] + 
                            q_{m2} \mbox{Tr} [N_2 \rho_{in}]    
                                                                   \\   \nonumber
& &\bar{F}_B^m = \frac{1}{4\pi} \int_0^{\pi} d\theta \int_0^{2\pi} d \phi \sin \theta
                             F_{B}^m (\theta, \phi).
\end{eqnarray}
When there is no noise, it is straightforward to show that $F_{B}^m (\theta, \phi)$ and 
$\bar{F}_B^m$ become
\begin{eqnarray}
\label{fidel-B-4}
& &F_{B}^{m=1} (\theta, \phi) = F_{B}^{m=2} (\theta, \phi) = 
\frac{1}{2} \left[1 + \cos^2 2 \nu \cos^2 \theta - \sin 2 \nu \cos 2 \nu \sin \theta 
                    \cos \theta \cos \phi \right]           \\   \nonumber
& &F_{B}^{m=3} (\theta, \phi) = F_{B}^{m=4} (\theta, \phi) = 
\frac{1}{2} \left[1 - \cos^2 2 \nu \cos^2 \theta + \sin 2 \nu \cos 2 \nu \sin \theta 
                    \cos \theta \cos \phi \right]           \\   \nonumber
& & \bar{F}_B^{m=1} = \bar{F}_B^{m=2} = \frac{2}{3} - \frac{1}{6} \sin^2 2 \nu
                                                           \\   \nonumber
& & \bar{F}_B^{m=3} = \bar{F}_B^{m=4} = \frac{1}{3} + \frac{1}{6} \sin^2 2 \nu.
\end{eqnarray}
When $m=1$ or $2$, $\bar{F}_B^m$ reaches to its maximum value $2/3$ if $\nu=0$ and $\nu=\pi/2$.
At the same time the Charlie's fidelity $\bar{F}_C$ becomes to its minimum value $2/3$.
When $\bar{F}_B^m$ reaches to its minimum value $1/2$ at $\nu=\pi/4$, $\bar{F}_C$ becomes
to its maximum value $1$. Thus, one can increase/decrease $\bar{F}_B^m$ at the cost of 
decreasing/increasing $\bar{F}_C$.

\begin{figure}[ht!]
\begin{center}
\includegraphics[height=10cm]{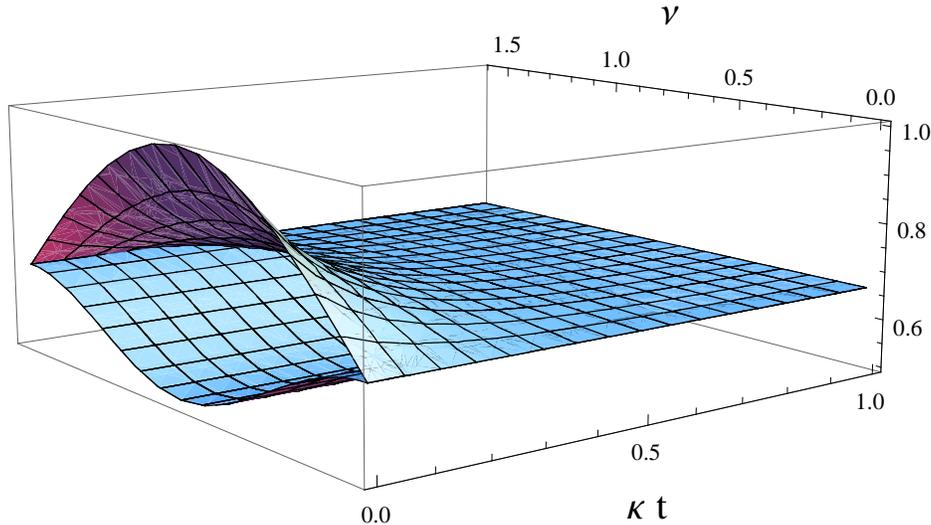}
\caption[fig1]{The $\nu$- and $\kappa t$-dependence  of $\bar{F}_C$ and 
$\bar{F}_B^m$ with $m=1, 2$ when the type of noise is $(L_{2,z}, L_{3,z}, L_{4,z})$. The upper
and lower surfaces correspond to $\bar{F}_C$ and $\bar{F}_B^m$ respectively. The difference 
between $\bar{F}_C$ and $\bar{F}_B^m$ is maximized in the $\kappa t \rightarrow 0$ limit. 
However, this difference becomes negligible with increasing $\kappa t$. This is due to
the fact that decoherence is a major dominant effect in the region of large $\kappa t$.}  
\end{center}
\end{figure}

The Bob's fidelities $F_B^m (\theta, \phi)$ and $\bar{F}_B^m$ are summarized in Table III when
the various noisy channels are introduced. One of the interesting points of Table III is that
the Bob's fidelities for the Z-noisy channel is independent of the noise parameter $\kappa$
while the Charlie's fidelities is dependent on $\kappa$ as Table II indicated. The $\nu$- and 
$\kappa t$-dependence of $\bar{F}_C$ and $\bar{F}_B^m (m=1,2)$ in $(L_{2,z}, L_{3,z}, L_{4,z})$
noisy channel is plotted together in Fig. 1. 
The upper surface in the figure corresponds to $\bar{F}_C$ 
and the lower one to $\bar{F}_B^m$. The difference between $\bar{F}_C$ and $\bar{F}_B^m$ is 
averagely maximized when $\kappa = 0$, which means there is no noise. If, however, $\kappa t$
becomes larger and larger, the difference between two fidelities becomes negligible. This is 
due to the fact that the effect of noise is significant compared to the choice of $\nu$ in the 
Bob's measurement. One can find a similar behaviors in the other noisy channels although we 
have not presented the $\nu$- and $\kappa t$-dependence of the fidelities explicitly in this 
paper.

\section{$\pi$-tangle}

In this section we will compute the $\pi$-tangle of the various noisy channels 
defined in Eq.(\ref{pi-1}). When there is no noise, it is easy to show that
\begin{equation}
\label{pi-11}
|| \rho_{GHZ}^{T_A} || = || \rho_{GHZ}^{T_B} || = || \rho_{GHZ}^{T_C} || = 2,
\end{equation}
which results in
\begin{equation}
\label{pi-12}
{\cal N}_{A(BC)} = {\cal N}_{B(AC)} = {\cal N}_{C(AB)} = 1.
\end{equation}
In addition, one can show that there is no contribution to the entanglement from the 
two-tangles in GHZ state:
\begin{equation}
\label{pi-13}
{\cal N}_{AB} = {\cal N}_{AC} = {\cal N}_{BC} = 0.
\end{equation}
Thus, $\pi$-tangle for the GHZ state is simply 
\begin{equation}
\label{pi-14}
\pi_{ABC}^{GHZ} = 1,
\end{equation}
which indicates that the GHZ state is a maximally entangled state.

\begin{center}
\begin{tabular}{c||c} \hline
Type of noise &   $\pi$-tangle              
                                                                      \\   \hline \hline
no noise & $1$                                                 \\   \hline
X noise &
$ e^{-8 \kappa t}$   \\    \hline
Y noise &
$\frac{1}{64} \bigg[ |1 - 3 e^{-2 \kappa t} - e^{-4 \kappa t} - e^{-6 \kappa t}| $  \\
{} &                                                   
 $- (1 - 3 e^{-2 \kappa t} - e^{-4 \kappa t} - e^{-6 \kappa t}) \bigg]^2$  \\ \hline
Z noise & $e^{-12 \kappa t}$                                                  
                                     \\   \hline
Isotropy &  $\frac{1}{64} \bigg[ |1 - e^{-8 \kappa t} - 4 e^{-12 \kappa t}|$  \\
noise & 
$ - (1 - e^{-8 \kappa t} - 4 e^{-12 \kappa t}) \bigg]^2$   \\ \hline

\end{tabular}

\vspace{0.1cm}
Table IV: The $\pi$-tangles for the various noisy channels
\end{center}
\vspace{0.5cm}

The $\pi$-tangles for $(L_{2,x}, L_{3,x}, L_{4,x})$, $(L_{2,y}, L_{3,y}, L_{4,y})$,
$(L_{2,z}, L_{3,z}, L_{4,z})$, and isotropy channels can be computed straightforwardly. 
For all noisy channels ${\cal N}_{A(BC)} = {\cal N}_{B(AC)} = {\cal N}_{C(AB)}$ and 
${\cal N}_{AB} = {\cal N}_{AC} = {\cal N}_{BC} = 0$ hold. This seems to be due to the fact that 
we considered only same-axis noisy channels. The $\pi$-tangles for the various noisy channels
are summarized at Table IV.  The interesting fact Table IV indicates is that while 
the $\pi$-tangles for the X- and Z-noise channels vanish at $\kappa t \rightarrow \infty$
limit, those for the Y- and isotropy-noise channels goes to zero at 
$y_* \leq \kappa t \leq \infty$ and $i_* \leq \kappa t \leq \infty$ respectively, where
\begin{eqnarray}
\label{pi-15}
& &y_* = \ln \frac{1 + (19 + 3 \sqrt{33})^{1/3} + (19 - 3 \sqrt{33})^{1/3}}{3}
       \sim 0.609378                                        \\   \nonumber
& &i_* = \frac{1}{4} \ln \frac{(54 + 3 \sqrt{321})^{1/3} + (54 - 3 \sqrt{321})^{1/3}}{3}
       \sim 0.146435.
\end{eqnarray}

\begin{figure}[ht!]
\begin{center}
\includegraphics[height=6cm]{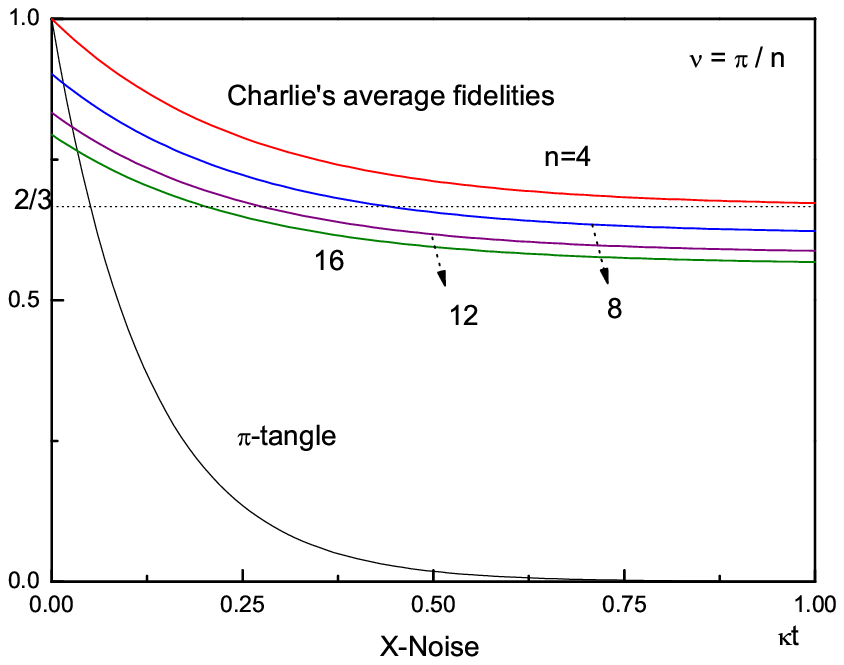}
\includegraphics[height=6cm]{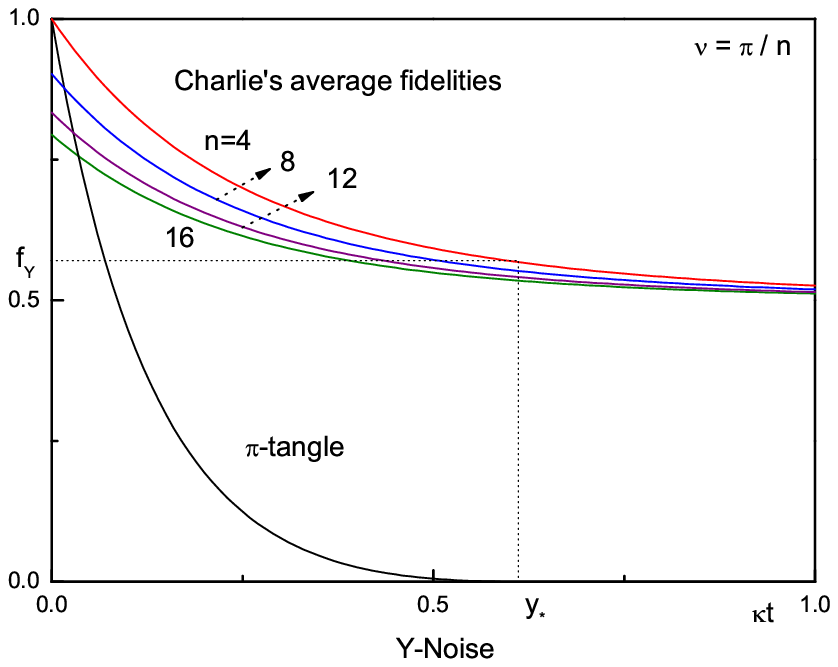}
\includegraphics[height=6cm]{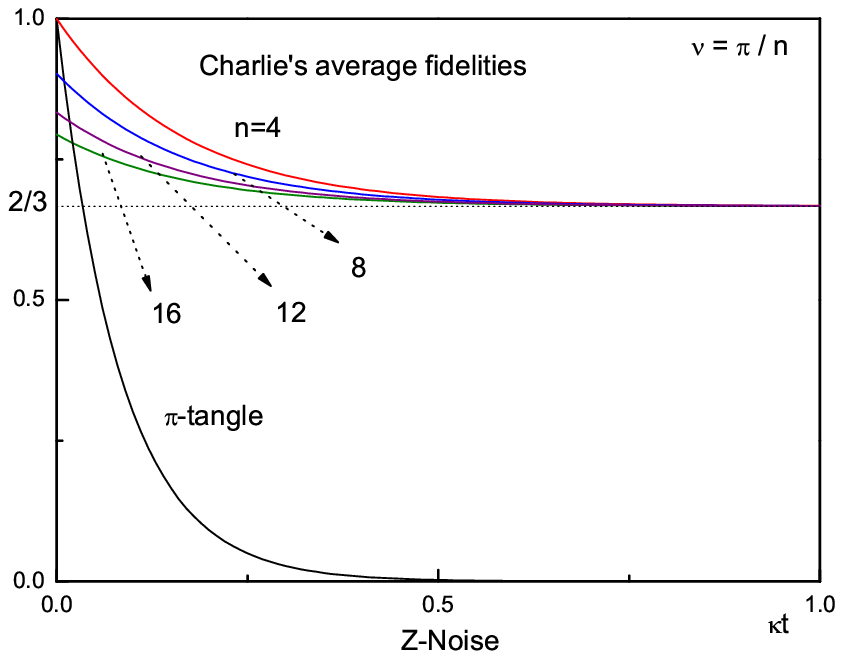}
\includegraphics[height=6cm]{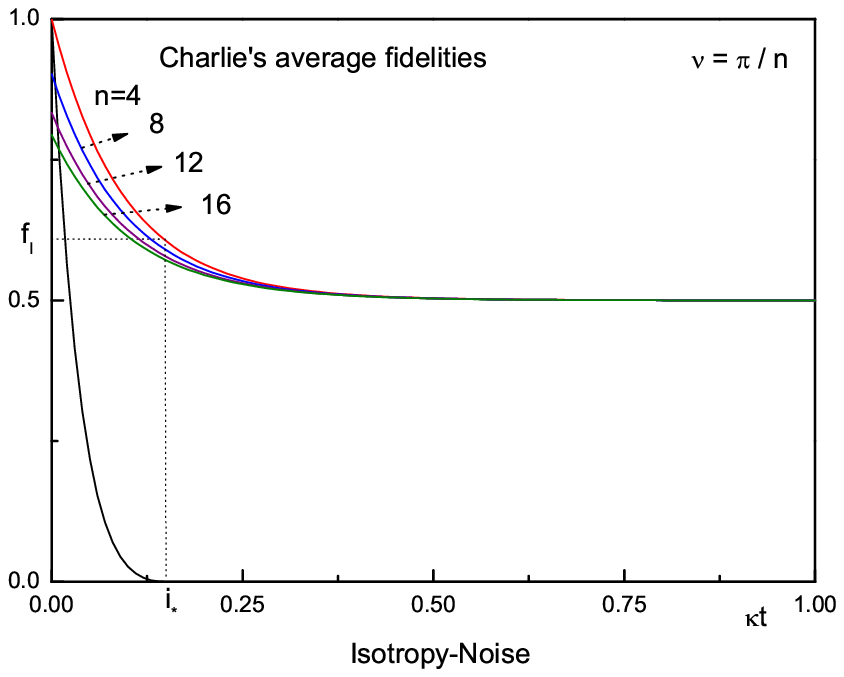}
\caption[fig2]{ The $\kappa t$ dependence of $\pi$-tangles and Charlie's average
fidelities $\bar{F}_C$ in $(L_{2,x}, L_{3,x}, L_{4,x})$ (Fig. 2a),
$(L_{2,y}, L_{3,y}, L_{4,y})$ (Fig. 2b), $(L_{2,z}, L_{3,z}, L_{4,z})$ (Fig. 2c), and
isotropy (Fig. 2d) noisy channels. }
\end{center}
\end{figure}

The $\kappa t$-dependence of $\pi$-tangles together with Charlie's average fidelity $\bar{F}_C$
for the various noisy channels are plotted in Fig. 2. In the Z-noisy channel the $\pi$-tangle
vanishes at $\kappa t = \infty$ and at this limit $\bar{F}_C$ goes to $2/3$ regardless of 
$\nu$, which is a classical fidelity limit. In the X-noise channel $\bar{F}_C$ goes to 
$(3 + \sin 2 \nu) / 6$ at $\kappa t = \infty$. When $\nu = \pi / 4$, this also goes to 
$2 /3$. Therefore, the $\pi$-tangles for X- and Z-noise channels seem to show a nice
connection between the Charlie's average fidelity and three-way entanglement of the
given channel.

However, this nice property is not maintained in the Y- and isotropy-noise channel. In the 
Y-noise channel the $\pi$-tangle vanishes at $y_* \leq \kappa t$. At $\kappa t= y_*$
the Charlie fidelity reduces to 
$ 0.166667 (3.08738 + 0.321426 \sin 2 \nu) $,
whose maximum is $f_Y = 0.568314$. Thus $f_Y$ is much less than the classical fidelity
limit $2/3$. Similar behavior can be found in the isotropy channel. In this channel the 
$\pi$-tangle vanishes at $i_* \leq \kappa t$. At $\kappa t = i_*$ the maximum Charlie's 
fidelity becomes $f_I = 0.609159$, which is also less than the classical limit $2/3$.

\section{three-tangle}
In this section we would like to discuss the three-tangles for the various noisy channels 
expressed in Eq.(\ref{summary1}).

\subsection{ $(L_{2,z}, L_{3,z}, L_{4,z})$ noisy channel}
Let us consider the pure state
\begin{equation}
\label{tangle-z-1}
|Z(z, \varphi)\rangle = \sqrt{z} |GHZ,1\rangle - e^{i \varphi} \sqrt{1-z} |GHZ,2\rangle
\end{equation}
where $z = (1 + e^{-6 \kappa t}) / 2$. It is easy to show that the three-tangle of 
$|Z(z, \varphi)\rangle$ is 
\begin{equation}
\label{tangle-z-2}
\tau_3 \left(|Z(z, \varphi)\rangle \right) = (1 - 2 z + 2 z^2) - 2 z (1-z) \cos 2\varphi.
\end{equation}
Thus, $\tau_3 \left(|Z(z, \varphi)\rangle \right)$ has a minimum at $\varphi=0$ and 
$\varphi=\pi$, i.e.
\begin{equation}
\label{tangle-z-3}
\tau_3 \left(|Z(z, 0)\rangle \right) = \tau_3 \left(|Z(z, \pi)\rangle \right) = (1 - 2 z)^2.
\end{equation}
In terms of the terminologies of Ref.\cite{oster07} $(1-2z)^2$ forms a convex characteristic
curve in $(z, \tau_3 \left(|Z(z, \varphi)\rangle \right))$ plane. In addition, one can show 
straightforwardly that $\varepsilon(\rho_{GHZ})$ defined in Eq.(\ref{summary1}) can be 
decomposed into
\begin{equation}
\label{tangle-z-4}
\varepsilon(\rho_{GHZ}) = \frac{1}{2} |Z(z, 0)\rangle \langle Z(z,0)| + \frac{1}{2}
|Z(z, \pi)\rangle \langle Z(z,\pi)|.
\end{equation}
If Eq.(\ref{tangle-z-4}) is optimal, then the three-tangle for $\varepsilon(\rho_{GHZ})$ is 
$(2z - 1)^2$. Since this coincides with the convex characteristic curve, Eq.(\ref{tangle-z-4})
should be the optimal decomposition. Thus, the three-tangle for the $\varepsilon(\rho_{GHZ})$
is 
\begin{equation}
\label{tangle-z-5}
\tau_{ABC}^z = (1 - 2 z)^2 = e^{-12 \kappa t}.
\end{equation}
It is interesting to note that the three-tangle and $\pi$-tangle are same with each other
in this channel.

\subsection{ $(L_{2,x}, L_{3,x}, L_{4,x})$ noisy channel}
Before we start computation, it is worthwhile noting that 
as shown in Ref.\cite{jung09-1} the state
\begin{equation}
\label{tangle-x-1}
\Pi_{GHZ} = \frac{1}{3} \bigg[ |GHZ,3\rangle \langle GHZ,3| + |GHZ,5\rangle \langle GHZ,5|
           + |GHZ,7\rangle \langle GHZ,7| \bigg]
\end{equation}
has vanishing three-tangle. This fact is shown in appendix A.

Now, let us consider a pure state
\begin{eqnarray}
\label{tangle-x-2}
& & |X(x, \varphi_1, \varphi_2, \varphi_3)\rangle = \sqrt{x} |GHZ,1\rangle - 
e^{i \varphi_1} \sqrt{\frac{1-x}{3}} |GHZ,3\rangle \\   \nonumber 
& & \hspace{3.0cm} -     
e^{i \varphi_2} \sqrt{\frac{1-x}{3}} |GHZ,5\rangle  -   
e^{i \varphi_3} \sqrt{\frac{1-x}{3}} |GHZ,7\rangle
\end{eqnarray}
where $x = (1 + 3 e^{-4\kappa t})/4$. Then it is easy to show that the three-tangle of 
$|X(x, \varphi_1, \varphi_2, \varphi_3)\rangle$ becomes
\begin{eqnarray}
\label{tangle-x-3}
& &\tau_3 \left(|X(x, \varphi_1, \varphi_2, \varphi_3)\rangle \right)   \\   \nonumber
& & = \Bigg| x^2 + \frac{(1-x)^2}{9} \left( e^{4i\varphi_1} + e^{4i\varphi_2} + 
                                           e^{4i\varphi_3} \right) -
\frac{2}{3} x (1-x) \left( e^{2i\varphi_1} + e^{2i\varphi_2} + e^{2i\varphi_3} \right)
                                                                        \\   \nonumber
& & \hspace{.5cm} - \frac{2}{9} (1-x)^2 \left( e^{2i (\varphi_1 + \varphi_2)} + 
e^{2i (\varphi_1 + \varphi_3)} + e^{2i (\varphi_2 + \varphi_3)}   \right) - 
\frac{8 \sqrt{3}}{9} \sqrt{x (1 - x)^3} e^{i (\varphi_1 + \varphi_2 + \varphi_3)}  \Bigg|.
\end{eqnarray}

The vectors $ |X(x, \varphi_1, \varphi_2, \varphi_3)\rangle$ has following properties. 
The three-tangle of it has the largest zero at $x=x_0 \equiv 3/4$ and 
$\varphi_1 = \varphi_2 = \varphi_3 = 0$. The vectors $|X(x, 0, 0, 0)\rangle$,
$|X(x, 0, \pi, \pi)\rangle$, $|X(x, \pi, 0, \pi)\rangle$ and $|X(x, \pi, \pi, 0)\rangle$ have
same three-tangles. Finally, $\varepsilon_X (\rho_{GHZ})$ can be decomposed into
\begin{eqnarray}
\label{tangle-x-4}
& &\varepsilon_X (\rho_{GHZ}) = \frac{1}{4} \Bigg[
|X(x, 0, 0, 0)\rangle \langle X(x, 0, 0, 0)| + 
|X(x, 0, \pi, \pi)\rangle \langle X(x, 0, \pi, \pi)|    \\  \nonumber
& & \hspace{2.0cm}
+ |X(x, \pi, 0, \pi)\rangle \langle X(x, \pi, 0, \pi)| + 
|X(x, \pi, \pi, 0)\rangle \langle X(x, \pi, \pi, 0)|   \Bigg].
\end{eqnarray}

When $x \leq x_0$, one can construct the optimal decomposition in the following form:
\begin{eqnarray}
\label{tangle-x-5}
& &\varepsilon_X (\rho_{GHZ}) = \frac{x}{4 x_0} \Bigg[
|X(x_0, 0, 0, 0)\rangle \langle X(x_0, 0, 0, 0)| + 
|X(x_0, 0, \pi, \pi)\rangle \langle X(x_0, 0, \pi, \pi)|    \\  \nonumber
& & \hspace{2.0cm}
+ |X(x_0, \pi, 0, \pi)\rangle \langle X(x_0, \pi, 0, \pi)| + 
|X(x_0, \pi, \pi, 0)\rangle \langle X(x_0, \pi, \pi, 0)|   \Bigg]  \\  \nonumber
& & \hspace{6.0cm}
+ \frac{x_0 - x}{x_0} \Pi_{GHZ}.
\end{eqnarray}
Since $\Pi_{GHZ}$ has the vanishing three-tangle, one can show easily 
\begin{equation}
\label{tangle-x-6}
\tau^X_{ABC} = 0  \hspace{1.0cm} \mbox{when} \hspace{,3cm} x \leq x_0 = 3/4.
\end{equation}

Now, let us consider the three-tangle of $\varepsilon_X (\rho_{GHZ})$ in the region
$x_0 \leq x \leq 1$. Since Eq.(\ref{tangle-x-4}) is an optimal decomposition at $x=x_0$,
one can conjecture that it is also optimal in the region $x_0 \leq x$. As will be shown
shortly, however, this is not true at the large-$x$ region. If we compute the three-tangle 
under the condition that Eq.(\ref{tangle-x-4}) is optimal at $x_0 \leq x$, its expression
becomes
\begin{equation}
\label{tangle-x-7}
\alpha_I^X (x) = x^2 - \frac{1}{3} (1 - x)^2 - 2 x (1 - x) - \frac{8 \sqrt{3}}{9}
\sqrt{x (1 - x)^3}.
\end{equation}
However, one can show straightforwardly that $\alpha_I^X (x)$ is not a convex function
in the region $x \geq x_*$, where
\begin{equation}
\label{tangle-x-8}
x_* = \frac{1}{4} \left( 1 + 2^{1/3} + 4^{1/3} \right) \approx 0.961831.
\end{equation}
Therefore, we need to convexify $\alpha_I^X (x)$ in the region $x_1 \leq x \leq 1$ to make 
the three-tangle to be convex function, where $x_1$ is some number between 
$x_0$ and $x_*$. The number $x_1$ will be determined shortly.

In the large $x$-region one can derive the optimal decomposition in a form:
\begin{eqnarray}
\label{tangle-x-9}
& &\varepsilon_X (\rho_{GHZ})    \\   \nonumber
& & = \frac{1-x}{4 (1-x_1)} \Bigg[
|X(x_1, 0, 0, 0)\rangle \langle X(x_1, 0, 0, 0)| +
|X(x_1, 0, \pi, \pi)\rangle \langle X(x_1, 0, \pi, \pi)|    \\  \nonumber
& & \hspace{2.5cm}
+ |X(x_1, \pi, 0, \pi)\rangle \langle X(x_1, \pi, 0, \pi)| +
|X(x_1, \pi, \pi, 0)\rangle \langle X(x_1, \pi, \pi, 0)|   \Bigg]  \\  \nonumber
& & \hspace{4.0cm}
+ \frac{x - x_1}{1 - x_1} |GHZ,1\rangle \langle GHZ,1|
\end{eqnarray}
which gives a three-tangle as 
\begin{equation}
\label{tangle-x-10}
\alpha_{II}^X (x, x_1) = \frac{1 - x}{1 - x_1} \alpha_I^X (x_1) + \frac{x - x_1}{1 - x_1}.
\end{equation}
Since $d^2 \alpha_{II}^X / dx^2 = 0$, there is no convex problem if $\alpha_{II}^X (x, x_1)$
is a three-tangle in the large-$x$ region.
The constant $x_1$ can be fixed from the condition of minimum $\alpha_{II}^X$, 
i.e. $\partial \alpha_{II}^X (x, x_1) / \partial x_1 = 0$, which gives
\begin{equation}
\label{tangle-x-11}
x_1 = \frac{1}{4} (2 + \sqrt{3}) \approx 0.933013.
\end{equation}
As expected $x_1$ is between $x_0$ and $x_*$. Thus, finally the three-tangle for 
$\varepsilon_X (\rho_{GHZ})$ becomes
\begin{eqnarray}
\label{tangle-x-12}
\tau_{ABC}^X = \left\{     \begin{array}{cc}
                        0   & \hspace{1.0cm}  x \leq x_0    \\
                        \alpha_I^X (x)    & \hspace{1.0cm}   x_0 \leq x \leq x_1    \\
                        \alpha_{II}^X (x, x_1)    & \hspace{1.0cm}    x_1 \leq x \leq 1
                           \end{array}                                    \right.
\end{eqnarray}
and the corresponding optimal decompositions are Eq.(\ref{tangle-x-5}), Eq.(\ref{tangle-x-4})
and Eq.(\ref{tangle-x-9}) respectively.
In terms of $\kappa t$ $\tau_{ABC}^X$ reduces to 
\begin{eqnarray}
\label{tangle-x-13}
\tau_{ABC}^X = \left\{    \begin{array}{cc}
                 \alpha_{II}^X (x, x_1)    & \hspace{1.0cm} 0 \leq \kappa t \leq \mu_1^X  \\
                 \alpha_I^X (x)    & \hspace{1.0cm} \mu_1^X \leq \kappa t \leq \mu_2^X \\
                 0                &  \hspace{1.0cm}  \mu_2^X \leq \kappa t \leq \infty
                          \end{array}                                                 \right.
\end{eqnarray}
where  $x = (1 + 3 e^{-4 \kappa t}) / 3$ and 
\begin{equation}
\label{tangle-x-14}
\mu_1^X = -\frac{1}{4} \ln \frac{4 x_1 - 1}{3} \approx 0.0233899  
\hspace{1.0cm}
\mu_2^X = - \frac{1}{4} \ln \frac{2}{3} \approx 0.101366.
\end{equation}

\subsection{$(L_{2,y}, L_{3,y}, L_{4,y}$) noisy channel}

The mixed state $\varepsilon_Y (\rho_{GHZ})$ given in Eq.(\ref{summary1}) can be re-written as
\begin{equation}
\label{tangle-y-1}
\varepsilon_Y (\rho_{GHZ}) = \xi \Pi_1^{GHZ} (Y_1) + (1 - \xi) \Pi_2^{GHZ} (Y_2)
\end{equation}
where 
\begin{eqnarray}
\label{tangle-y-2}
& &\Pi_1^{GHZ} (Y_1) = Y_1 |GHZ,1\rangle \langle GHZ,1| \\  \nonumber
& &  \hspace{2.0cm}
+ \frac{1 - Y_1}{3} \bigg[ |GHZ,3\rangle \langle GHZ,3| + |GHZ,5\rangle \langle GHZ,5| + 
  |GHZ,7\rangle \langle GHZ,7| \bigg]
                                                                     \\   \nonumber
& &\Pi_2^{GHZ} (Y_2) = Y_2 |GHZ,2\rangle \langle GHZ,2|   \\  \nonumber
& &  \hspace{2.0cm}
+ \frac{1 - Y_2}{3} \bigg[ |GHZ,4\rangle \langle GHZ,4| + |GHZ,6\rangle \langle GHZ,6| + 
  |GHZ,8\rangle \langle GHZ,8| \bigg].
\end{eqnarray}
In Eq.(\ref{tangle-y-2}) the constants are given by
\begin{equation}
\label{tangle-y-3}
\xi = \frac{y_+ (y_+^2 + 3 y_-^2)}{8}     \hspace{1.0cm}
Y_1 = \frac{y_+^2}{y_+^2 + 3 y_-^2}       \hspace{1.0cm}
Y_2 = \frac{y_-^2}{3 y_+^2 + y_-^2}
\end{equation}
where $y_{\pm} = 1 \pm e^{-2 \kappa t}$. It is worthwhile noting that $\Pi_2^{GHZ} (Y_2)$ is 
local-unitary (LU) equivalent to $\Pi_1^{GHZ} (Y_2)$, i.e. 
$$ \Pi_1^{GHZ} (Y_2) = (\sigma_z \otimes \openone \otimes \openone) \Pi_2^{GHZ} (Y_2)
(\sigma_z \otimes \openone \otimes \openone)^{\dagger}. $$
Since the three-tangle is LU-invariant quantity, the three-tangle for $\Pi_2^{GHZ} (Y_2)$ should
be equal to that for $\Pi_1^{GHZ} (Y_2)$. Since $\Pi_1^{GHZ} (Y_2)$ can be obtained from
$\varepsilon_X (\rho_{GHZ})$ by replacing $x$ by $Y_2$, one can compute the three-tangle for
$\Pi_2^{GHZ} (Y_2)$ directly from Eq.(\ref{tangle-x-12}). Since, furthermore, $Y_2 \leq 1/4$
in the entire range of $\kappa t$, the three-tangle for $\Pi_2^{GHZ} (Y_2)$ should be 
zero.  

Since $\varepsilon_Y(\rho_{GHZ})$ is rank-$8$ mixed state, it seems to be highly difficult to 
compute the three-tangle. Still we do not know how to compute it analytically. However, one can
compute its upper bound as following. Since the three-tangle for $\Pi_2^{GHZ} (Y_2)$ is zero
and the three-tangle for the mixed state is obtained by the convex-roof method, 
Eq.(\ref{tangle-y-1}) implies that the three-tangle for $\varepsilon_Y(\rho_{GHZ})$ should be less
than $\xi$ times three-tangle for $\Pi_1^{GHZ} (Y_1)$. Since $\Pi_1^{GHZ} (Y_1)$ is same with
$\varepsilon_X(\rho_{GHZ})$ if $x$ is replaced by $Y_1$, one can compute the upper bound of 
the three-tangle for $\varepsilon_Y(\rho_{GHZ})$, $\tau_{ABC}^{Y:UB}$ directly from 
Eq.(\ref{tangle-x-12}). The superscript UB stands for upper bound. The final result of this 
upper bound can be summarized as
\begin{eqnarray}
\label{tangle-y-4}
\tau_{ABC}^{Y:UB} = \left\{     \begin{array}{cc}
                \xi \alpha_{II}^X (Y_1, x_1)  & \hspace{1.0cm} 0 \leq \kappa t \leq \nu_1^*  \\
                \xi \alpha_I^X (Y_1)    & \hspace{1.0cm} \nu_1^* \leq \kappa t \leq \nu_2^*  \\
                0                       & \hspace{1.0cm} \nu_2^* \leq \kappa t \leq \infty
                                 \end{array}                   \right.
\end{eqnarray}
where
\begin{equation}
\label{tangle-y-5}
\nu_1^* = -\frac{1}{2} \ln (\sqrt{3} - 1) \sim 0.155953   \hspace{1.0cm}
\nu_2^* = \frac{1}{2} \ln 2 \sim 0.346574.
\end{equation}
Of course, $x_1$ is given in Eq.(\ref{tangle-x-11}).

\subsection{isotropy noisy channel}

The mixed state $\varepsilon_I (\rho_{GHZ})$ given in Eq.(\ref{summary1}) can be re-written as
\begin{equation}
\label{tangle-i-1}
\varepsilon_I (\rho_{GHZ}) = \zeta \Sigma_1^{GHZ} + (1 - \zeta) \Sigma_2^{GHZ}
\end{equation}
where 
\begin{eqnarray}
\label{tangle-i-2}
& &\Sigma_1^{GHZ} = \left(\frac{1}{2} + \frac{2 p^3}{1 + 3 p^2} \right) 
                                 |GHZ,1\rangle \langle GHZ,1|           \\   \nonumber
& &\hspace{3.0cm} + 
\left(\frac{1}{2} - \frac{2 p^3}{1 + 3 p^2} \right) |GHZ,2\rangle \langle GHZ,2|   
                                                            \\   \nonumber
& &\Sigma_2^{GHZ} = \frac{1}{6} \left\{ I - \left(|000\rangle \langle 000| + |111\rangle
                                                  \langle 111| \right) \right\}
\end{eqnarray}
and
\begin{equation}
\label{tangle-i-3}
\zeta = \frac{1 + 3 p^2}{4}
\end{equation}
with $p = e^{-4 \kappa t}$.

The state $\varepsilon_I (\rho_{GHZ})$ is rank-8 mixed state and we do not know how to 
compute the three-tangle of it exactly. Since, however, the three-tangle of $\Sigma_2^{GHZ}$
is zero, one can compute at least the upper bound as $\zeta$ times three-tangle of 
$\Sigma_1^{GHZ}$. This upper bound can be easily computed by making use of the analytical 
result of the three-tangle for the Z-noise channel.
The final result of this upper bound is 
\begin{equation}
\label{tangle-i-4}
\tau_{ABC}^{I:UB} = \frac{4 p^6}{1 + 3 p^2} = \frac{4 e^{-24 \kappa t}}{1 + 3 e^{-8 \kappa t}}
\end{equation}
where the superscript UB stands for the upper bound.

\section{Conclusion}

\begin{figure}[ht!]
\begin{center}
\includegraphics[height=6cm]{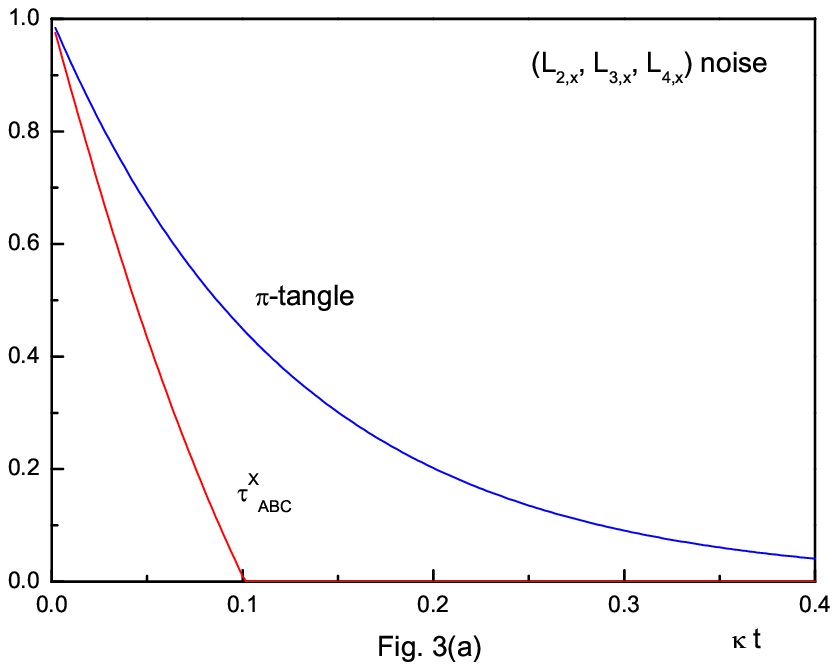}
\includegraphics[height=6cm]{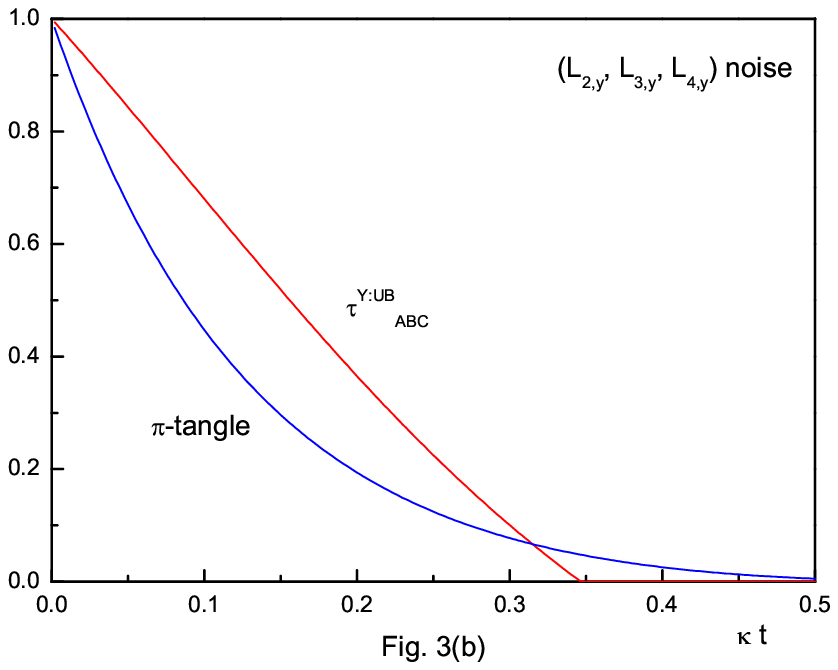}
\includegraphics[height=6cm]{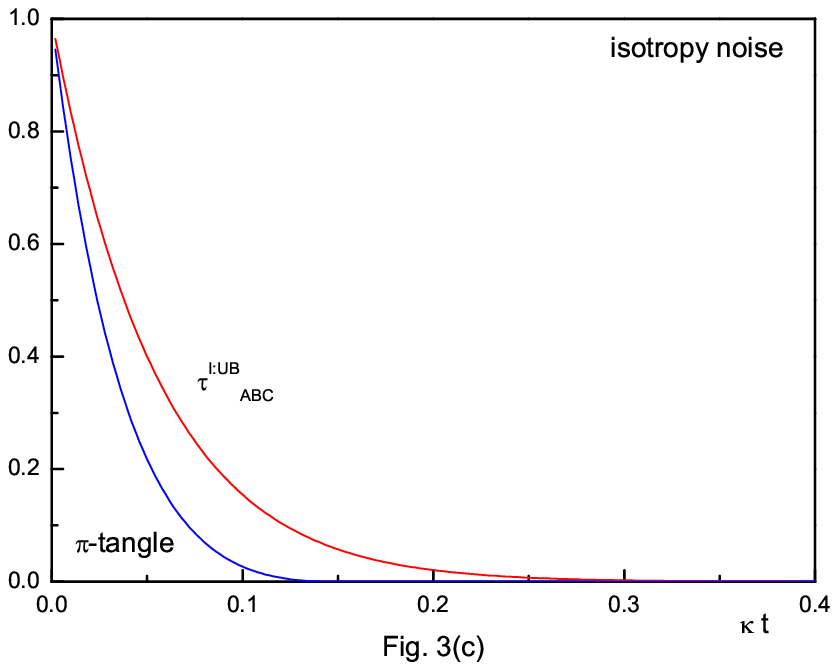}
\caption[fig3]{ The $\kappa t$ dependence of three-tangle and $\pi$-tangle for  
$(L_{2,x}, L_{3,x}, L_{4,x})$ (Fig. 3a),
$(L_{2,y}, L_{3,y}, L_{4,y})$ (Fig. 3b), and 
isotropy (Fig. 3c) noisy channels. }
\end{center}
\end{figure}

In this paper we computed the $\pi$-tangles explicitly for the mixed states summarized in 
Eq.(\ref{summary1}). It is shown that the $\pi$-tangles for the X- and Z-noisy channels
vanish at $\kappa t \rightarrow \infty$, where the maximum Charlie's fidelities reduce to
the classical limit $2/3$. However, this nice property is not maintained for Y- and 
isotropy-noise channels. For Y-noise the $\pi$-tangle vanishes at $y_* \leq \kappa t$, where
$ y_*$ is given at Eq.(\ref{pi-15}). At $\kappa t = y_*$ the maximum Charlie's fidelity becomes 
$0.57$, which is much less than the classical limit. For isotropy noise the $\pi$-tangle 
vanishes at $i_* \leq \kappa t$. At $\kappa t = i_*$ the maximum Charlie's fidelity 
becomes $0.61$, which is also less than the classical limit. Although the $\pi$-tangle was 
constructed in Ref.\cite{ou07-1} to reflect the three-party entanglement of the W-type 
states, it does not seem to give a meaningful interpretation in the real quantum information 
process.

We also computed the three-tangles for the X- and Z-noise channels. The remarkable fact is that 
the three-tangle for the Z-noise channel is exactly same with the corresponding 
$\pi$-tangle. Therefore, the three-tangle for the Z-noise channel vanishes at 
$\kappa t \rightarrow \infty$, where all Charlie's fidelities reduce to the classical limit
regardless of Bob's measurement outcome. For X-noise the $\kappa t$-dependence of the 
three tangle is plotted in Fig. 3(a). For comparison we plotted the corresponding $\pi$-tangle
together. As Fig. 3(a) has shown, the three-tangle is much less than the corresponding
$\pi$-tangle. In this channel the three-tangle vanishes at $\mu_2^X \leq \kappa t$, where
$\mu_2^X = -(1/4) \ln (2/3)$. At $\kappa t = \mu_2^X$ the Charlie's fidelity becomes 
$(11 + 5 \sin 2 \nu) / 18$. When, therefore, $\nu = (1/2) \sin^{-1} (1/5) \sim 0.100679$,
Charlie's fidelity reduces to the classical limit $2/3$. However, the maximum Charlie's fidelity
goes to $8/9$, which is much larger that the classical limit. 

The $\kappa t$-dependence of 
$\tau^{Y:UB}_{ABC}$ and $\tau^{I:UB}_{ABC}$ are plotted in Fig. 3(b) and Fig. 3(c) respectively.
For comparison we plotted the corresponding $\pi$-tangle together. Fig. 3(b) shows that 
$\tau^{Y:UB}_{ABC}$ is larger than the corresponding $\pi$-tangle at 
$0 \leq \kappa t \leq 0.315$. Fig. 3(c) shows that $\tau^{I:UB}_{ABC}$ is larger than the 
corresponding $\pi$-tangle in the entire range of $\kappa t$. This is due to the fact that 
$\tau^{Y:UB}_{ABC}$ and $\tau^{I:UB}_{ABC}$ are merely the upper bounds of the real 
three-tangles for Y- and isotropy-noise channels. If the calculational tool for the 
three-tangle of the arbitrary three-party mixed states are developed someday, 
the real three-tangles computed via this tool should be less than the corresponding 
$\pi$-tangles.

In this paper we examined the physical meaning of the three-tangle and $\pi$-tangle 
in the real quantum information process. We adopted the three-party teleportation via 
various noisy channels as a model of quantum process.
It is shown that the $\pi$-tangle seems to be too large to have a meaningful interpretation.
Although we cannot compute the three-tangles for Y- and isotropy-noise channels due to their
high rank, the results for X- and Z-noise seems to imply the fact that the three-tangle 
is too small to have meaningful interpretation. Probably we need a different three-party
entanglement measure whose value is between three-tangle and $\pi$-tangle.

{\bf Acknowledgement}: 
This work was supported by the Kyungnam University
Foundation Grant, 2008.

\newpage

\begin{appendix}{\centerline{\bf Appendix A}}

\setcounter{equation}{0}
\renewcommand{\theequation}{A.\arabic{equation}}

In this appendix we would like to prove that $\Pi_{GHZ}$ defined in Eq.(\ref{tangle-x-1}) has
vanishing three-tangle. Consider a pure state
\begin{equation}
\label{appen-1}
|J(\theta_1, \theta_2)\rangle = \frac{1}{\sqrt{3}} |GHZ,3\rangle - \frac{1}{\sqrt{3}}
e^{i \theta_1} |GHZ,5\rangle - \frac{1}{\sqrt{3}} e^{i \theta_2} |GHZ,7\rangle.
\end{equation}
Then, it is easy to show that the three-tangle of $|J(\theta_1, \theta_2)\rangle$ is 
\begin{equation}
\label{appen-2}
\tau_3(\theta_1, \theta_2) = \frac{1}{9} |1 - \left(e^{i \theta_1} - e^{i \theta_2} \right)^2 |
                     |1 - \left(e^{i \theta_1} + e^{i \theta_2} \right)^2 |,
\end{equation}
which vanishes when
\begin{eqnarray}
\label{appen-3}
& &(i) \hspace{.5cm} e^{i \theta_1} - e^{i \theta_2} = 1 \Longrightarrow 
(\theta_1 = \pi/3, \theta_2=2\pi/3),
                       (\theta_1 = 5\pi/3, \theta_2=4\pi/3)  
                                                                  \\  \nonumber
& &(ii) \hspace{.5cm} e^{i \theta_1} - e^{i \theta_2} = -1 \Longrightarrow 
(\theta_1 = 2\pi/3, \theta_2=\pi/3),
                       (\theta_1 = 4\pi/3, \theta_2=5\pi/3)  
                                                                  \\  \nonumber
& &(iii) \hspace{.5cm} e^{i \theta_1} + e^{i \theta_2} = 1 \Longrightarrow 
(\theta_1 = \pi/3, \theta_2=5\pi/3),
                       (\theta_1 = 5\pi/3, \theta_2=\pi/3)  
                                                                  \\  \nonumber
& &(iv) \hspace{.5cm} e^{i \theta_1} + e^{i \theta_2} = -1 \Longrightarrow 
(\theta_1 = 2\pi/3, \theta_2=4\pi/3),
                       (\theta_1 = 4\pi/3, \theta_2= 2\pi/3).
\end{eqnarray} 
Furthermore, one can show straightforwardly that $\Pi_{GHZ}$ can be decomposed into
\begin{eqnarray}
\label{appen-4}
& &\Pi_{GHZ} = \frac{1}{8} \Bigg[ |J\left(\frac{\pi}{3}, \frac{2 \pi}{3} \right) \rangle
\langle J\left(\frac{\pi}{3}, \frac{2 \pi}{3} \right) | + 
|J\left(\frac{\pi}{3}, \frac{5 \pi}{3} \right) \rangle
\langle J\left(\frac{\pi}{3}, \frac{5 \pi}{3} \right) | 
                                                         \\  \nonumber
& & \hspace{2.0cm} + 
|J\left(\frac{2\pi}{3}, \frac{\pi}{3} \right) \rangle
\langle J\left(\frac{2\pi}{3}, \frac{\pi}{3} \right) | +
|J\left(\frac{2\pi}{3}, \frac{4 \pi}{3} \right) \rangle
\langle J\left(\frac{2\pi}{3}, \frac{4 \pi}{3} \right) | 
                                                          \\   \nonumber
& & \hspace{2.0cm} + 
|J\left(\frac{4\pi}{3}, \frac{2 \pi}{3} \right) \rangle
\langle J\left(\frac{4\pi}{3}, \frac{2 \pi}{3} \right) | + 
|J\left(\frac{4\pi}{3}, \frac{5 \pi}{3} \right) \rangle
\langle J\left(\frac{4\pi}{3}, \frac{5 \pi}{3} \right) | 
                                                         \\   \nonumber
& & \hspace{2.0cm} + 
|J\left(\frac{5\pi}{3}, \frac{ \pi}{3} \right) \rangle
\langle J\left(\frac{5\pi}{3}, \frac{\pi}{3} \right) | + 
|J\left(\frac{5\pi}{3}, \frac{4 \pi}{3} \right) \rangle
\langle J\left(\frac{5\pi}{3}, \frac{4 \pi}{3} \right) |    \Bigg]. 
\end{eqnarray}
Combining Eq.(\ref{appen-3}) and (\ref{appen-4}), one can show that Eq.(\ref{appen-4}) is 
the optimal decomposition of $\Pi_{GHZ}$ and the three-tangle is  
\begin{equation}
\label{appen-5}
\tau_3 \left(\Pi_{GHZ}\right) = 0.
\end{equation}
\end{appendix}

\end{document}